# Stigmergy in Comparative Settlement Choice and Palaeoenvironment Simulation

Eugene Ch'ng
*School of Computer Science*
*University of Nottingham Ningbo China*
*199 Taikang East Road, Zhejiang, Ningbo 315100*
eugene.chng@nottingham.edu.cn

Vince Gaffney and Gido Hakvoort
*Digital Humanities Hub*
*The University of Birmingham*
*Edgbaston B15 2TT, United Kingdom*
v.l.gaffney@bham.ac.uk and
gido.hakvoort@gmail.com

**Abstract**

Decisions on settlement location in the face of climate change and coastal inundation may have resulted in success, survival or even catastrophic failure for early settlers in many parts of the world. In this study we investigate various questions related to how individuals respond to a palaeoenvironmental simulation, on an interactive tabletop device where participants have the opportunity to build a settlement on a coastal landscape, balancing safety and access to resources, including sea and terrestrial foodstuffs, whilst taking into consideration the threat of rising sea levels. The results of the study were analysed to consider whether decisions on settlement were predicated to be near to locations where previous structures were located, stigmergically, and whether later settler choice would fare better, and score higher, as time progressed. The proximity of settlements was investigated and the reasons for clustering were considered. The interactive simulation was exhibited to thousands of visitors at the 2012 Royal Society Summer Science Exhibition at the "Europe's Lost World" exhibit. 347 participants contributed to the simulation, providing a sufficiently large sample of data for analysis.

**Keywords:** *stigmergy, virtual environments, agent-based modelling, simulation, archaeology*

## 1. Introduction

Global warming at the end of the last Ice Age led to the inundation of vast landscapes that had once been home to thousands of people. These lost lands represent one of the last frontiers of geographical and archaeological exploration and exist off the coast of Europe, the Gulf and in the South China Sea and the Sunda Shelf. Whilst the inundated landscape cannot be explored conventionally, pioneering work using seismic reflectance data has provided the first detailed maps of the prehistoric topography and the results have received international recognition as being globally important with respect of our understanding of these enigmatic landscapes but also in respect of our understanding of past and present response to rising sea levels and climate change (http://e-a-a.org/EHP_2013.pdf). Following the publication of these unique datasets from the area of the North Sea (also known as Doggerland), there has been an opportunity for further reconstruction of the landscape, environment, flora and fauna, through GIS and environmental modelling, and occasionally this has involved the use of interactive virtual environments (Ch'ng, 2007b; Ch'ng and Stone 2006; Fitch, 2013; Gaffney, Fitch, & Smith, 2009). More recently agent-based modelling and simulation have emerged as technologies that might also provide insights into the changing ecology of Doggerland and related landscapes in the Gulf and Asia. The potential of such applications to add to the debate follows from the increasing availability of relevant data and this suggests that these technologies may provide a step-change to our understanding of these complex, historic environments (Benjamin, 2011; Ch'ng & Gaffney, 2013; Gaffney *et al.*, 2009; Gaffney, Thomson, & Fitch, 2007; Kohler & Varien, 2012; Mithen, 2003). The distributed computational modelling of individual agents and complex ecological interactions between millions of agents and the environment will fill gaps in our knowledge that may be the result of the highly inaccessible nature of these lost landscapes (Gaffney *et al.*, 2013) and, significantly, such studies will permit the generation of different hypotheses and scenarios to be explored through dynamic simulations in a manner that has never previously been imagined by archaeological researchers (Craenen, Murgatroyd, Theodoropoulos, Gaffney, & Suryanarayanan, 2012; Kohler & Gumerman, 2000; Lake, 2013; Murgatroyd, Craenen, Theodoropoulos, Gaffney, & Haldon, 2012). Whilst complexity science-based archaeological simulation that uses human input has never been attempted before, a similar work by Goldstone and Roberts on self-organisation in trail systems in groups of humans within a virtual environment have demonstrated the potentials of such work for future studies (Goldstone & Roberts, 2006).

The "Europe's Lost World" group (Wickham-Jones, 2012), associated with such landscape research, received an invitation to present at the Royal Society's Summer Science Exhibition held at Carlton House Terrace, London, during the summer of 2012 (2-7 July). The exhibition incorporated an interactive tabletop simulation designed to investigate human decisions when presented with threats including rising tides linked to climate change and the need to balance requirements to maximise the opportunities provided by access to terrestrial and ecological resources required for subsistence.

The research sought to understand comparisons between modern responses and potential past behaviour and to what extent any decisions were balanced in respect of safety and basic necessities, using a large tabletop device as the medium between the simulated past and the present. The study also considered whether stigmergy would apply in such situations (Grassé, 1959; Grassé, 1984; Holland & Melhuish, 1999), i.e., would future settlers learn from previous inhabitants from the traces of past occupation left behind on the landscape. Furthermore, we investigated the clustering of settlements around strategic regions of the landscape in relation to their survival 'scores' and investigated the phenomena through statistical analysis. It is likely that this novel approach of exploring landscape use by merging virtual reconstructions with human interaction via a tabletop medium may provide us with insights into behavioural processes underpinning how early settlers interacted with past landscapes and their perception of earlier traces of occupation.

The article begins with the background to our research followed by a methodological section covering the reconstruction of a virtual environment and agent-based model that incorporates real data from previous studies (Ch'ng & Gaffney, 2013; Ch'ng & Stone, 2006; Gaffney *et al.*, 2009), using an interactive tabletop as the medium. Section three explains our approach to data collection on the settlement behaviours of 350 participants. Section four presents the statistical analysis of our results and finally the article concludes with a discussion and outlines opportunities for future work.

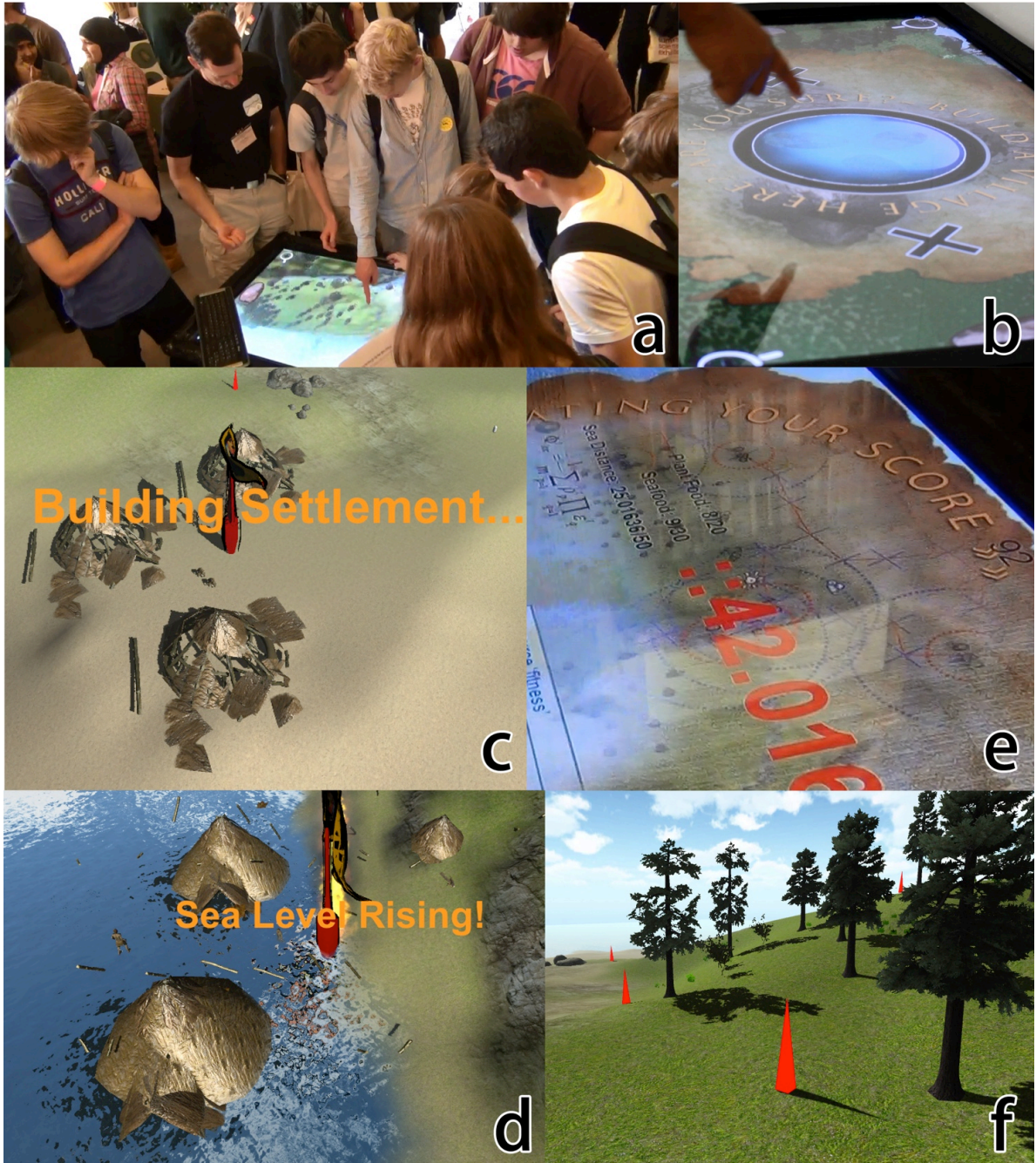

*Figure 1. The Palaeoenvironment Simulation: a) participants interacting with the simulation on a tabletop computer, b) Tabletop User Interface confirming the building of settlement, c) Mesolithic people building a settlement, d) Sea-level rouse, destroying the settlement. e) Scores calculated from the simulation, weighing between safety from sea level rises and resources. f) Pyramidal 3D icons in the landscape are previous settlement locations.*

## 2. Background

The issues of stigmergy are rarely considered within archaeology but may be an important factor when researching the complexities of settlement or past behaviour. Within Hunter-Gatherer studies, for instance, researchers often provide substantive information on the nature of mobility and the extent of territories that may be exploited by individual groups. It is probably true that single hunter-gatherers territories may have been very extensive in the past and individual groups certainly ranged across the land according to need and the availability of seasonal resources. However, in some societies the territories themselves may also shift cyclically over a longer period (Binford and Cherry 1983, Chapter 6, Fig. 52). Consequently, landscapes may appear empty to new settlers or migrants although they may well have been an active part of a seasonally occupied territory or one that may have been revisited periodically as part of a longer behavioural pattern. In these instances the proxy traces of past activity may well have been discernable to groups entering the landscape for the first time. These situations could have engendered locales that attracted persistent occupation. This may have been particularly important during periods, such as the early Holocene, when marine inundation may have forced hunter-gatherer groups from traditional lands and into the seasonally occupied territories of other groups. How such incomers explored or interpreted the traces of past occupation is generally never considered by archaeologists whose evidence is rarely that fine-grained. However, sites with access to specific resources (and perhaps identified from cultural markers) may have been the scenes of violent confrontation during these periods as competing groups sought to control land and resources (Mithen 2003, p.175). It is also notable that at specific periods hunter-gatherers may have begun to construct other structures including houses that may have been markers for important locales as much as places to live in (Gaffney *et al.*, 2009; Waddington & Bailey, 2007).

There is another context in which we might begin to identify the effects of stigmergy as a past cultural process and that relates to the evolution of ritual and the creation of monuments during the Mesolithic period. The emergence of ritual activity and the origins and evolution of monuments are extremely problematic areas of research within archaeology. It is clear from the study of later monuments, including exceptional sites such as Stonehenge, that, despite the apparent complexity of such structures, there is no original plan for their development and that they evolve over very long periods of time. An example of a very early monument studied by the authors demonstrates this. The site at Warren Fields in Aberdeenshire (Scotland) has been interpreted as a primitive time reckoner dated toward the beginning of the $8^{th}$ millennium BC (Gaffney *et al.*, 2013). A series of 12 pits, up to 2, 3 metres in width and spread over c. 90 metres, seem to mimic the phases of the moon and may have been used to track time, month by month, following the moon's phases. To ensure that the asynchronicity of the lunar month and solar year did not impact on seasonal integrity, the structure was also aligned to provide an annual solar correction on the winter solstice. However, whilst the highly distinctive format of the pits, from waxing, gibbous, full through the waning phases of the moon, appears to suggest a cohesive and planned monument the structure actually emerged over a period of hundreds of years. There is no evidence that this could have been planned and suggests that an original astronomic observation, and repeated visits to the site, allowed the organic growth of a cohesive structure with a currently unique time-keeping function. This is all the more striking when it is appreciated that the dating for the site suggests that the pits were still being marked when early farming settlements arrived some 4000 years later. There is no suggestion that these communities were related but this does suggest that the stigmergic effect of this unique monument was still active over millennia of human settlement and that this may well have been reflected in the decision to build a very rare Neolithic timber hall a very short distance from the pre-existing, Mesolithic monument (Murray, Murray, & Fraser, 2009).

The experience of colonising a totally unknown or wholly unoccupied landscape must have happened at some point globally, yet such events, historically, are extremely rare. Re-settlement, possibly into landscapes which may appear relatively under-utilised, are more frequently attested throughout human history. However, the experience of such settlement across the tapestry of human history is rarely well documented even within historic periods. This becomes yet more complex during remote periods, such as the resettlement of Europe after the last glaciation, when textual evidence is absent and even the topography of the land may well have changed significantly along with the climate, floral and faunal environment. In such circumstances the paucity of direct evidence for settlement, and in the case of early Holocene inundated landscapes, access to primary evidence may be limited or impossible. In such instances researchers must explore other methodological avenues to try to understand the behavioural processes associated with settlement expansion and contraction and, perhaps, ultimately to use such information to inform future exploration in a directed manner. In respect of re-settlement the concept of stigmergy may provide some insight into past behavioural practise. Stigmergy, first coined by Pierre Paul Grassé (Grassé, 1959) refers to the effect of pre-existing environmental states on the actions of termites building a mound. Grasse noted that "the coordination of tasks and the regulation of constructions do not depend directly on workers, but the construction themselves. The worker does not direct his work, he is guided by it…" Studies in stigmergy usually relate to social insects (Bonabeau, 1999; Goss, Beckers, Deneubourg *et al.*, 1990; Theraulaz & Bonabeau, 1999) and how they coordinate via stimulus-response mechanism for coordinated behaviour. The term is simply defined as "coordination without direct communication" which social insects such as ants, bees and wasps depend on for survival as a collective organism. It would appear that at an individual level each insect works in solitude yet, at the collective level, an apparent work of coordination is observed. Some insects depended on pheromone trails left behind by other insects; others rely on information gathered from the environment for progressing the construction of nests. Whilst the term was applied in simple insect behaviour, stigmergy also occurs in highly intelligent species, including human beings (Helbing, Keltsch, & Molnar, 1997; Helbing, Schweitzer, Keltsch, & Molnár, 1997), and work is being carried out on facilitating such cooperation in crowd-sourced cuneiform fragment reconstructions (Ch'ng, Lewis, Gehlken, Woolley, 2013).

With this in mind the interactive tabletop simulation of human settlement presented at the Royal Society provided data that could be used to study such behaviour. Analysis of the outputs was specifically guided by the requirement to understand whether the behaviour of participants was influenced by traces of earlier participants and whether this contributed to better performances by later players looking for new settlement locations. Three key theoretical concepts support this goal. The first is *interaction* (Rabaud, 1937; Theraulaz & Bonabeau, 1999) which posits that an individual's behaviour will act as a stimulus for modifying the behaviour of another. The other is *interattraction*, in which animal within a social species is attracted by other animal of the same species. The third is the *group effect* through which each individual is a source of stimuli for other individuals creating a diffusion of effect when a critical number of specific stimuli occur. Whilst our hypothesis suggests that evidence of the first two behaviours may be found in our analysis of the datasets, the group effect cannot occur, as the 347 participants did not build settlements simultaneously. Stigmergy as a mechanism, is based on self-organising processes, which in this study stands on Halley and Winkler's definition, it is "an interplay of internal and external sources of order" (Halley & Winkler, 2008).

## 3. Methods: Virtual Environment and Agent-based Model and Simulation

For these purposes we reconstructed a theoretical landscape that incorporates models of real-world data from previous studies (Ch'ng & Gaffney, 2013; Ch'ng & Stone, 2006; Gaffney *et al.*, 2009). The virtual palaeoenvironment is set during the Early Holocene c. 8,000 to 7,500BP. This period, also known as the Early Mesolithic in Europe is a climatic optimum. The period was represented as a maximum of temperatures, and evidence suggests it was warmer than average for the interglacial. During the period following the last Ice Age C. 18,000 BC, melting ice sheets ultimately submerged nearly half of Western Europe, creating bays and inlets along the Atlantic coast, which provided a new, rich ecosystem that was attractive to human occupation prior to these lands being lost to the sea as a consequence of sea level rise. Wetter and warmer, the period associated with the simulation may have been associated with a mixed forest of Oak, Elm, Common Lime (Linden), and Elder that spread northward, together with Pine and Hazel as the climate ameliorated. The simulation terrain reflects such features with special attention on hazel bushes which dot the open areas of the landscape as these are known to be an important resource during the period (Mithen, 2001). The coast has marine resources, including shellfish, also know as a key foodstuff from the archaeological record, and these are more abundant in specific areas, i.e., bays and inlets. This section describes briefly the technical details of the simulation that are relevant to the mechanics of our methodology. Highly technical details for the reconstruction of virtual environments, agent-based models, and interactive tabletop applications are beyond the scope of this paper but can be referenced from various sources from the authors (Ch'ng & Gaffney, 2013; Ch'ng, 2013).

### 3.1. A Interactive Tabletop for Navigating Ancient Landscapes

The advent of devices with touch-based natural user interfaces (Seow *et al.*, 2009) have revolutionised the way in which users access information. These touch-based displays have pervaded both workflows and leisurely activities for the majority of modern computer users, or indeed the general public using mobile devices with similar interfaces. The choice of a touch-based interactive tabletop display for answering our research questions is, we believe, appropriate for a number of reasons. Initially, the touch interface is now commonplace, and perhaps even expected in many sections of contemporary society. Such an interface removes the barrier associated with a computer mouse, or even an Xbox 360 controller. The horizontal interface allows a larger portion of the virtual palaeolandscape to be explored in a much shorter time than a conventional "first-person" experience of a virtual environment. This invariably decreases the required length of participation time and thus increases the opportunity for data sampling. Finally, a table configuration allows participants and audience to gather around in a familiar setting, i.e., around a table. Consequently, a tabletop displays allows co-located collaboration within a truly social setting (Shen *et al.*, 2006), which meets our goal of investigating potential stigmergic learning and co-located observations concerning virtual environments and agent-based modelling. It is also important to stress that whilst the tabletop display makes it possible for modern participants to participate in the creation of simulated ancient settlements, it is, in the end, a medium through which we can explore a poorly understood, but important, social process - stigmergy.

### 3.2. Virtual Environments

The visualisation aspects of the terrain and all dynamic objects within the virtual environment are modelled, textured and animated in Blender 3D before being ported into the Unity3D engine IDE where autonomous behaviour scripts (C# programming language) were embedded. Dynamic objects are items that have a life cycle within the simulation, as opposed to landscape features. These objects are autonomous, or can be interacted with by participants or other agents, i.e., Mesolithic human agents, Hazel bushes & nuts, house, village, seafood resources, etc. The simulation is event-based, a participants' interaction with the user interface trigger events related to the agent, causing them to trigger further events within the simulation until the cycle is completed and the final score calculated process. The events can be categorised into three types:

System and User Interface Events:
- Participants trigger events through the user interface which supports starting the simulation, navigating the landscape, building settlements, selecting confidence levels, username input for storing the simulation score, and ending the game.
- The planting of the flagpole, which participants use for navigating the landscape, triggers the virtual camera to navigate to the point on the landscape where the pole is planted.
- Timer events notify and warn participants of any imposed time limits, they also trigger sea level rise.

Environmental Events:
- The sea level, as part of the environment, is a dynamic object and has a simple script that causes it to rise when triggered at the end of the settlement build event.

Dynamic Object Events:
- All agents in the environment have events associated with them.
- Hazel bushes produce hazel nuts according to an internal timer event.
- Seafood gizmos dotting the coasts notify human agents of the presence of marine resource when they are nearby.
- Human agents roam the landscapes and build settlements when participants confirm the locations. They also gauge the scores and provide a report at the score calculation user interface at the end of the game.
- Mesolithic houses trigger the transition of the built process visualised in different stages. The completion of the village settlement triggers the sea level rise, which tests the coastal proximity of the settlement.

### 3.3. Agents and Scoring Mechanics

The agents – dynamic objects in the environment - are simplified interactive versions of recent developments (Ch'ng & Gaffney, 2013). The simplified version has fewer states and transitions than the autonomous version used for hypotheses testing in large simulations. States in a complex simulation of human agents in our recent work are "exploratory", "hunger", "tiredness", "dying", etc., and the associated transitions between each state. States for a hazel bush or an animal agent are birth, growth, reproduction, senescence and death. The mechanisms of agents in this simulation are simplified because they are partially autonomous, requiring inputs from human participants such as where to build a settlement. Terrestrial resource (forest food, wood, etc.) and marine resource (seafood) gizmos do not go through state transitions, they merely emit information about their location and resource information for agents to read. Mesolithic houses go through various state transitions as they are being built for visualisation purposes.

The scoring mechanism uses coastal proximity, access to marine and terrestrial resource as important elements in the "survivability" of a settlement. Coastal proximity has a different significance in the simulation in comparison to marine resource. Coastal proximity in the simulation can have two consequences. Negatively, being too near the coast means that the settlement will be in danger of flooding. Positively, being near the coast will mean that coastal resources will be more accessible. Coastal proximity includes sources where fresh water can be found within inlets and streams. On the other hand, coastal proximity does not necessarily mean that settlements will have access to marine resource, Marine resource are, again, differentially located in strategic features of a landscape, including inlets, rivers and rocky areas. Terrestrial resource includes land mammals, and hazel nuts from shrubs, which are usually located in open landscapes. The simulation places importance on the relative coastal proximity and marine resource.

The scoring mechanism depends on the location of the settlement identified by a participant and the available marine $\alpha$ and terrestrial resources $\beta$ in the vicinity, within range $r$ of agent $i$ at time $t$ in the simulation using the Iverson bracket,

$$\alpha = \sum_{j}^{m} \left[ \sqrt{(x_j^t - x_i^t)^2 + (y_j^t - y_i^t)^2} < r_{ji} \right] \tag{1}$$

$$\beta = \sum_{k}^{n} \left[ \sqrt{(x_k^t - x_i^t)^2 + (y_k^t - y_i^t)^2} < r_{ki} \right] \tag{2}$$

Where $j$ and $k$ are individual resource.

The scoring formula for coastal proximity is,

$$\gamma = \left| z_i^t - W^t \right| \tag{3}$$

Where $z$ is the height of agent $j$ in the settlement and $W$ is the sea level.

The fitness of the settlement measured with the Adaptability Measure (Ch'ng, 2007a) using the standard for coastal proximity and lower bound for marine and terrestrial resource. The total score (Eq. 4) composed of the three measures (Eq. 5-7) is therefore,

$$T = \alpha_{eff} + \beta_{eff} + \gamma_{eff} \tag{4}$$

$$\alpha_{eff} = f_{lower}\left(\alpha, L_{ji}, P_{ji}, U_{ji}, c_{ji}\right) w_\alpha \tag{5}$$

$$\beta_{eff} = f_{lower}\left(\beta, L_{ki}, P_{ki}, U_{ki}, c_{ki}\right) w_\beta \tag{6}$$

$$\gamma_{eff} = f\left(\gamma, L_\gamma, P_\gamma, U_\gamma, c_\gamma\right) w_\gamma \tag{7}$$

The weightings $w_\alpha + w_\beta + w_\gamma = 100$ posits the score based on the importance of the resource and environment factor towards the survivability of the settlement. In the simulation presented at the exhibition, the variables in Table 1 were used.

*Table 1. Variables for Equations 1 to 7:*

| Parameters | Values |
|---|---|
| $r_{ji}$ | 97 |
| $r_{ki}$ | 110 |
| $L_{ji}$ and $L_{ki}$ | 5 |
| $P_{ji}$ and $P_{ki}$ | 25 |
| $U_{ji}$ and $U_{ki}$ | 35 |
| $L_\gamma$ | 15 |
| $P_\gamma$ | 27 |
| $U_\gamma$ | 35 |
| $c_{ji}$, $c_{ki}$, $c_\gamma$ | 0.5 |
| $w_\alpha$ | 30 |
| $w_\beta$ | 20 |
| $w_\gamma$ | 50 |

## 3.4. Experimental Setup and Data Collection

The tabletop interactive simulation was exhibited to thousands of visitors on the "Europe's Lost World" stand at the 2012 Royal Society Summer Science Exhibition. Our tabletop exhibit was always crowded, peaking at lunchtime until the evening. At the end of the exhibition, 347 participants contributed to the simulation, leaving behind a large sample of datasets for research. The simulation ran in three sessions from 3 July 10:45am – 6 July 8:50pm. Session 1: 3 July 10:45am – 5 July 11:08am, Session 2: 5 July 11:26am – 6 July 3:53pm and Session 3: 6 July 4:01pm – 6 July 8:50pm. The experiment is not controlled, participants of all gender and age (8 to 65) participated freely, but those who used the simulation had a clear interest in the Royal Society's state-of-the-art exhibits. The mean time difference between simulation users was 7.05 minutes over all the sessions and including all 347 users. The simulation ran for a maximum of 3 minutes, the additional 4 minutes on average was used for resetting the simulation, explaining the concept of the simulation to the next user and viewing of a pre-simulation user interface tutorial. Most participants did not find the tabletop interaction difficult to learn, we believe that the learning curve was minimised due to the touch-based interaction similar to pervasive devices such as smartphones and tablet computers.

Participants went through a simple process in the simulation.
1. The purpose of the simulation and the background of the study were explained to the participants prior to the start of the experiment and included information on the prehistoric diet and life style, dangers of coastal inundation, etc. Explaining the imperative of survival to the participants prepared them for 'survival' (knowledge of survival strategies would have been passed from generation to generation in prehistory). Care was taken not to reveal strategic locations in the landscape. Participants were made aware of the experiment being undertaken and videoing was authorised by participants.
2. Prior to starting the simulation, all previous locations of participants of the session were loaded and a small red, pyramidal 3D icon (see figure 1) placed on the landscape to indicate previous settlements to the current participant. This information was intended to support the study of stigmergy.
3. Participants navigated and explored the landscape for the best location to build a Mesolithic settlement. The navigation widget is large pole with a flag that participants used for planting at a location which triggers the virtual camera to navigate to the destination where the flag is placed.
4. Participants selected the 'build' option and confirmed the settlement location with a confidence level indicator (participants indicate how confident they are with regards to their chosen settlement location)
5. Participants watched as agents built the settlement
6. A visualisation of sea level rise occurs (if the sea level rises above the settlement, a physics simulation of the destruction of the village is triggered)
7. At the end of the sea level rise, the user interface shows the calculation of the scores
8. Participants input their names.
9. All information (location of settlement as x, y, z, the confidence level, total score, marine and terrestrial resource scores, time used in seconds, a timestamp consisting of the year, month, day, hour, minutes and seconds) is recorded in a database.

The object of the simulation was to gather data that allowed researchers to begin to compare modern human decision-making with prehistoric behaviour, and to draw conclusions based on the presumption that when faced with the need to survive, participants using the simulation might generate similar behaviour to earlier settlers. We looked for evidence of learning and ascertained whether stigmergy works in such situations, i.e., future settlers read information left behind by past dwellers on the landscape and build there. We expected participants, who understood the importance of survival, to seek for strategic locations in situations of urgency rather than choosing less prosaic factors, i.e., the view from a specific location. To create a sense of urgency for our participants, we limited the simulation to 180 seconds. This imparts a sense of 'virtual' urgency to the simulation. Participants were notified when they had 60 seconds left before the sea level rose.

We understood that participants had only a limited time at the exhibit. We therefore replicated the mobility of prehistoric settlers, who might roam far and wide before settling down, with a top-down view of the landscape, on the tabletop display. This allowed them to explore much more of the landscape in a far shorter time than would their prehistoric predecessors.

## 4. Data Analysis

Here we investigated the data collected from participant interactions with the simulation over a span of 4 days. We first looked at the distribution of settlements across the landscape. Figure 2 depicts landscape features within the hypothetical environment that our participants explored. We deliberately made the island small to create a sense of urgency. We expected that having a small island also reduce participant exploration time, which increases of our chance to capture more data from more participants, thus increasing our sample size. There were only two small rivers within the small island and they were intended as a source of portable water. None of the participants were told about the best location to build a settlement at on the map. The participants were only told generally about the properties of food, and where hazels shrubs would grow.

### 4.1. Settlement Clusters and Heat Maps

Figure 3 shows the heat map and the sequence of settlements built by participants (1, 2, …, $S$) in three separate but continuous sessions. The Figure on the top left (a) shows the concentration of settlements as a heat map together with scores (0 to 100) overlaid on the map. The other figures (b to c) are data from individual sessions. Session 01 had 121 participants, session 02 had 39 participants, and session 03 had 187 participants. The heat map for each of the session and their settlement scores (The Total Score in Equation 4 – coastal proximity + marine resource + terrestrial resource) is colour coded in three ranges, 0-30, 31-60, and 61-100. Comparison of the heat maps (see figure guide lines) indicates that similar patterns emerge as the number of participants increases (Session 01 and 02), and that 'settlement builders' have a natural attraction toward certain features of the terrain. We would expect a similar pattern if the number of participants is increased in session 03. Participants' preference seems to follow a certain priority, such as the large river mouth, sparse woodlands near coastal areas and the lake. Our settlement builders kept away from islands, open areas including grasslands and coasts, ravines, treacherous coastal areas and rocky bays.

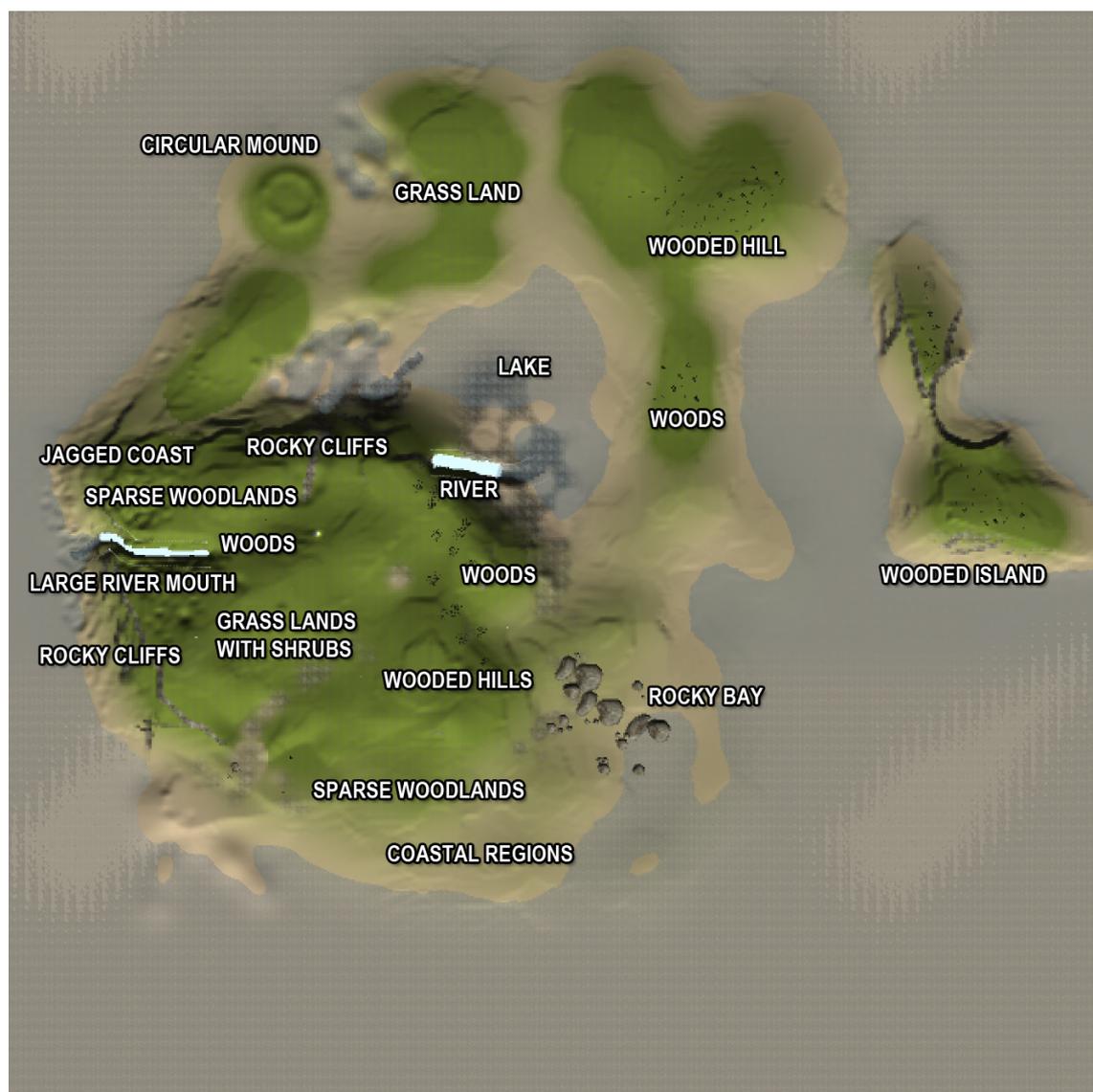

*Figure 2. A hypothetical landscape constructed as a virtual environment illustrating features of the terrain. Sources of fresh water are abundant around the landscape with inlets and streams that are not visible here.*

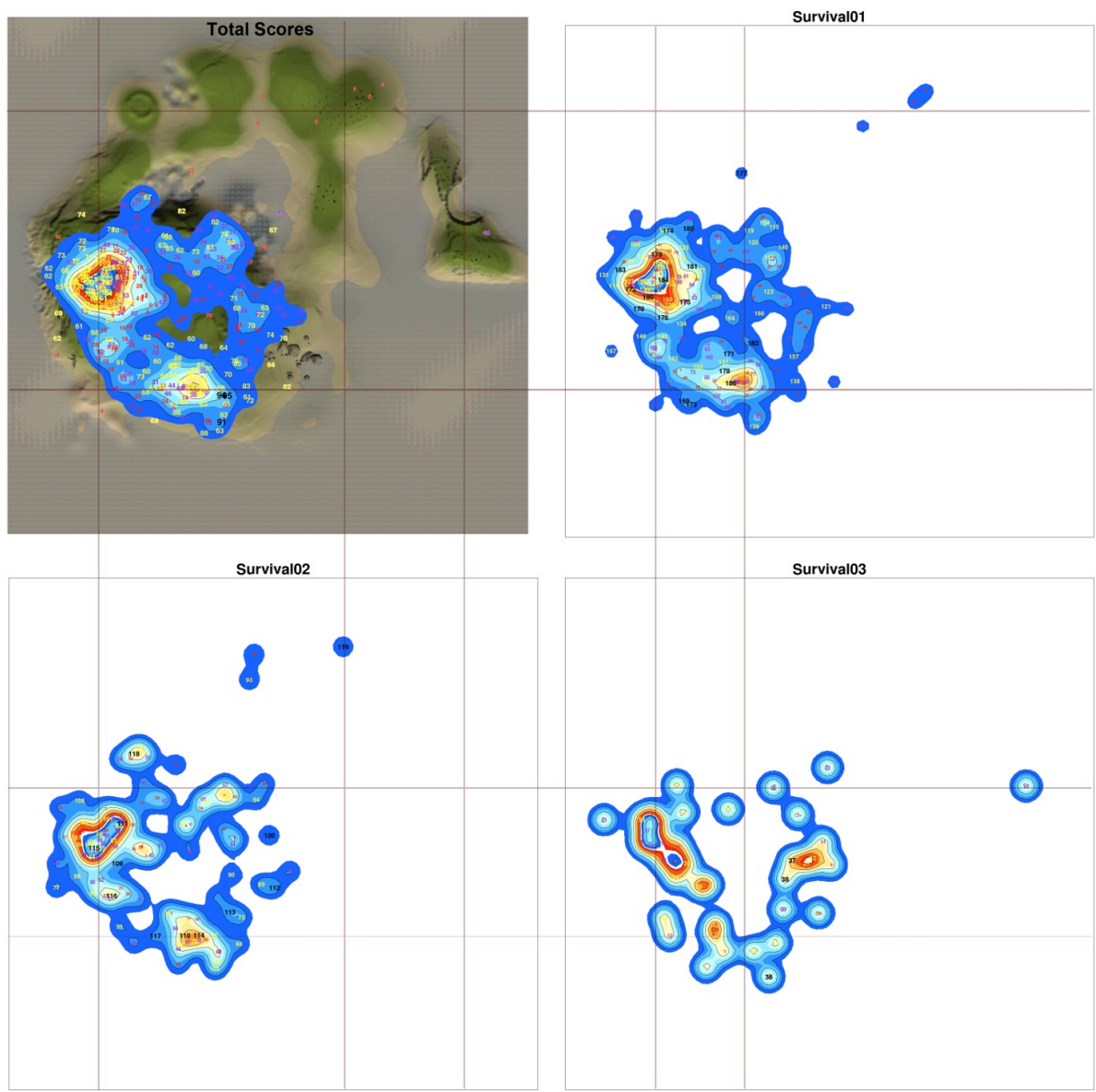

*Figure 3. Combined simulation scores (top left) of sessions Survival01 to Survival03 over four days of data collection at the Royal Society Summer Science Exhibition 2012. The maps on the top right, bottom right and bottom left show the sequence of participation indicated by numbers (0 is the first settler) overlaid on a heat map of settlements chosen by participants. Guidelines have been superimposed to aid in comparison of xy coordinates of the heat maps.*

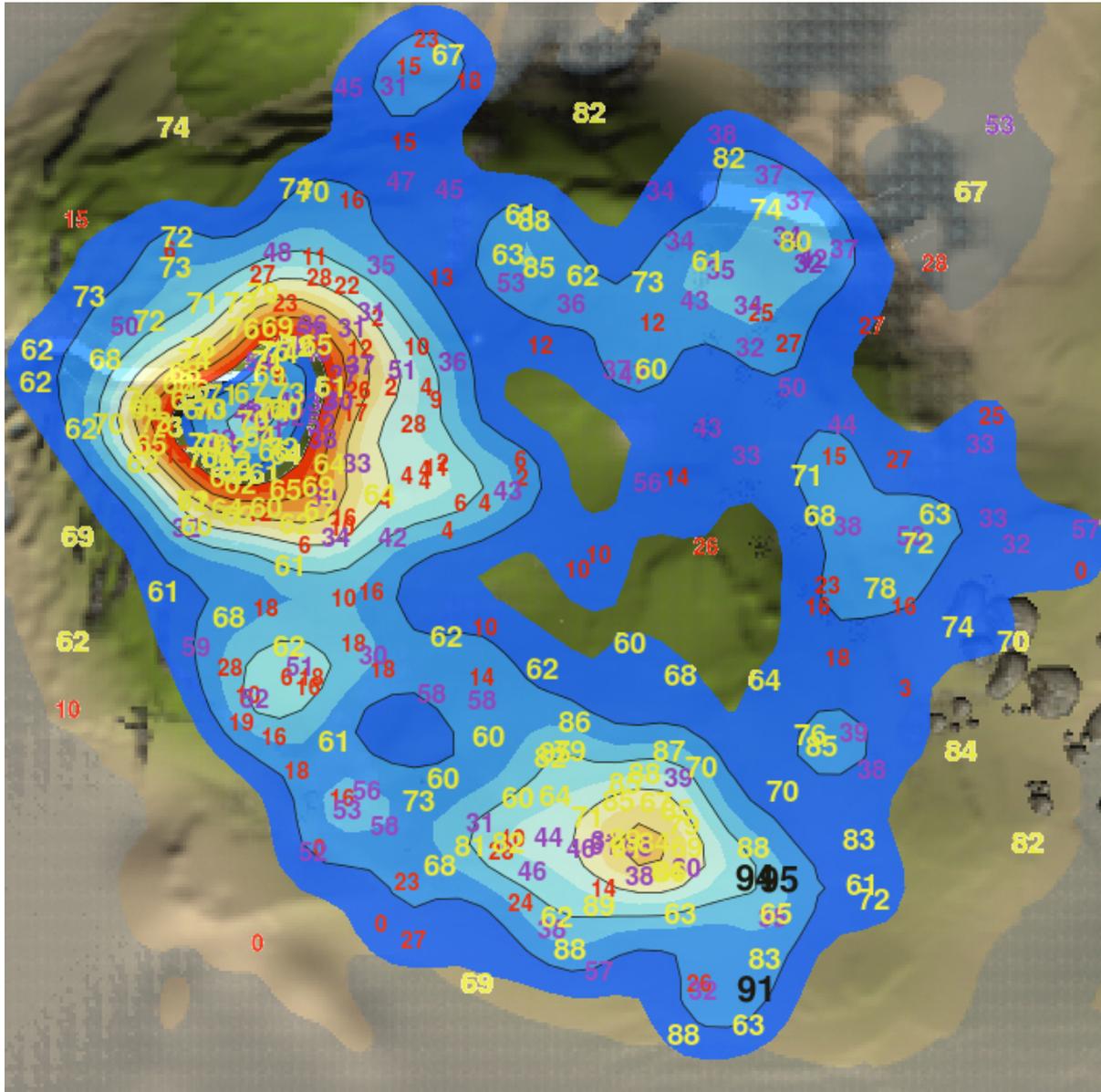

*Figure 4. Total scores for all sessions. Total Scores (Equation 4 – coastal proximity + marine resource + terrestrial resource) ranged from 0-100 overlaid on a heat map of settlements chosen by participants. Colour codes for the scores are used for clarity. High scores (in black, ranging from 91-100) are located at the lower right quadrant. Settlement positions of low scores (between 0 to 30 and in red) seem to be distributed quite evenly in the valleys. Intermediate scores (31-60, purple) have a pattern similar to the lower scoring groups. Scores 61-90 (yellow) show clustering and have a higher density in the areas where populations are more abundant.*

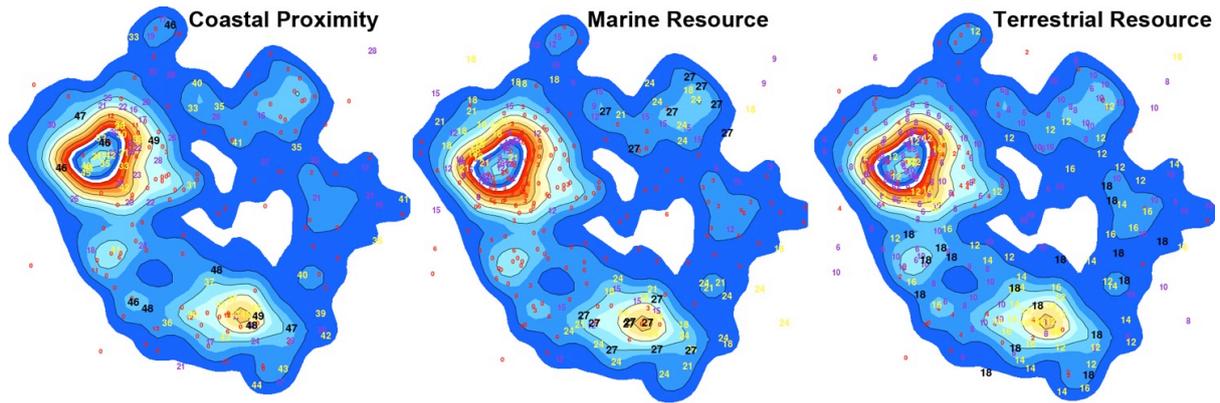

*Figure 5. Each score presented as a separate heat map, revealing further information on the scoring aspects of combined sessions (01+02+03). Coastal proximity scores (weighted 50% for Equation 3, therefore rates 0-50, 50 being the highest score for coastal proximity) of settlements. Marine resource scores (weighted 30% for Equation 1, range from 0-30) of settlements. Terrestrial resource scores (weighted 20% for Equation 2, 0-20) of settlements.*

### 4.2. Evidence of Stigmergy

In figure 6 we illustrate the normalised average scores for participant 1 to *n* for each session (01 to 03). It was expected that scores for the sessions would stabilise after a certain number of participants. Extreme scores would have a decreasing impact on the overall average. However, it should still be informative to see how the average scores progressed over time.

The first thing visible from the graphs in figure 6 is that scores in all three sessions stabilised roughly around their average values (a score of 0.5 multiplied by their maximum value). For sessions 1 and 2 this occurred approximately after the 100$^{th}$ participant, looking at the trend, session 3 is projected to follow a similar pattern. Secondly from the graphs we can see that the total score is strongly related to the coastal proximity score as it follows a similar pattern, this is largely due to coastal proximity's 0.5 weighted importance.

Looking at the individual scores it can be seen that participants tried to find a balance between coastal proximity and marine resources. If they were to build too close to the coast, their settlements would be flooded, conversely, if they build a settlement too far away, marine resources would have been out of reach and fresh water difficult to find. In session 1 participants settled areas too far from the coast, in session 3 participants' settlements were too close. In session 2 participants started off more in the lower central part of the island, too far away from the sea for marine resources, fresh water, and too low for survival against the rising tides.

After the initial batch of participants, no further improvements or declines in score were observed. This indicates that our participants either congregate (whether intentionally or not) around earlier settlements which had an average score initially, or, it could be that participants attempted to seek for extreme values. This resulted in both high and low scores and an average total score. Evidence suggests that the latter is a more likely phenomenon - given the strong and steep slopes at the beginning of the graphs. However, looking at the heat map in figure 4 we do see patterns of congregation. Using traditional k-means clustering (with *k=10*) we identified likely locations where participants might congregate. After discarding clusters with small number of participants (cluster size < 30), an average total score was calculated for each cluster, which is shown in figure 7. The majority of our participants (83%) built their settlements in one of these clusters, with an average total score of the clusters in the range [25,70]. This suggests that participants either worked collectively (simultaneously), or via stigmergy, by locating exit markers to identify best locations to build even though there were alternate locations which might also provide a 'good' place to build. This appears to follow the observation of an attractive force from the formation of an increasingly richer stimulatory environment via the appearance of individual settlers (Theraulaz, Bonabeau, & Deneubourg, 1998). The density of settlement near the sparse woodlands where a water source were, is, we believe, the work of stigmergy, where an initial clustering of settlers led to an attraction of more participants to the area. The initial attraction could be the result of the portable water, which is near the coastal areas where food is abundant and is in sparse woodlands where hazel shrubs grew. Further analysis below suggests that stigmergy was in operation during the experiment rather than simultaneous collective intelligence.

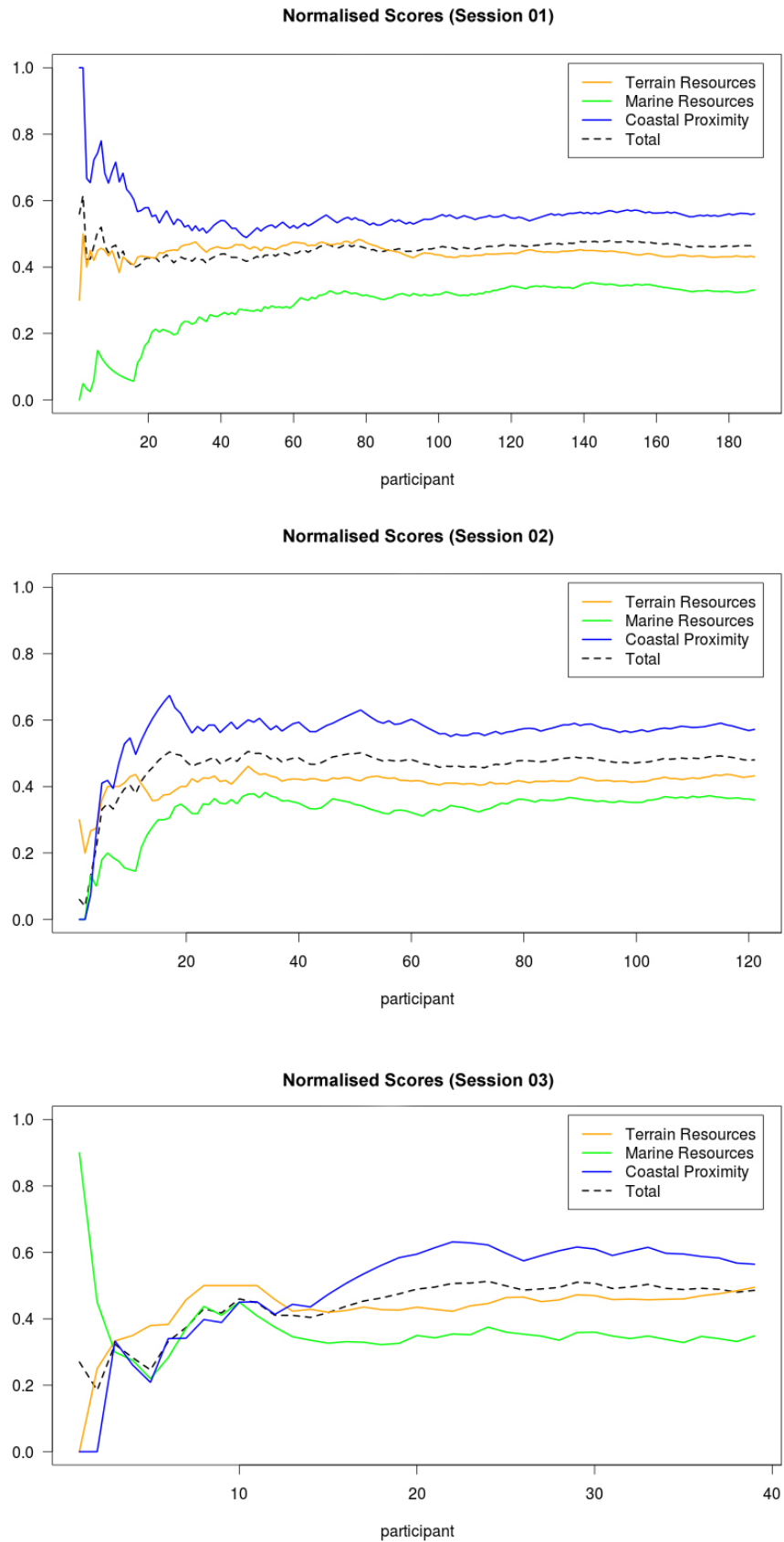

*Figure 6. Normalised average scores for participants 1 to n for each session (01 to 03). Scores stabilised around their average values (a score of 0.5 multiplied by their maximum value).*

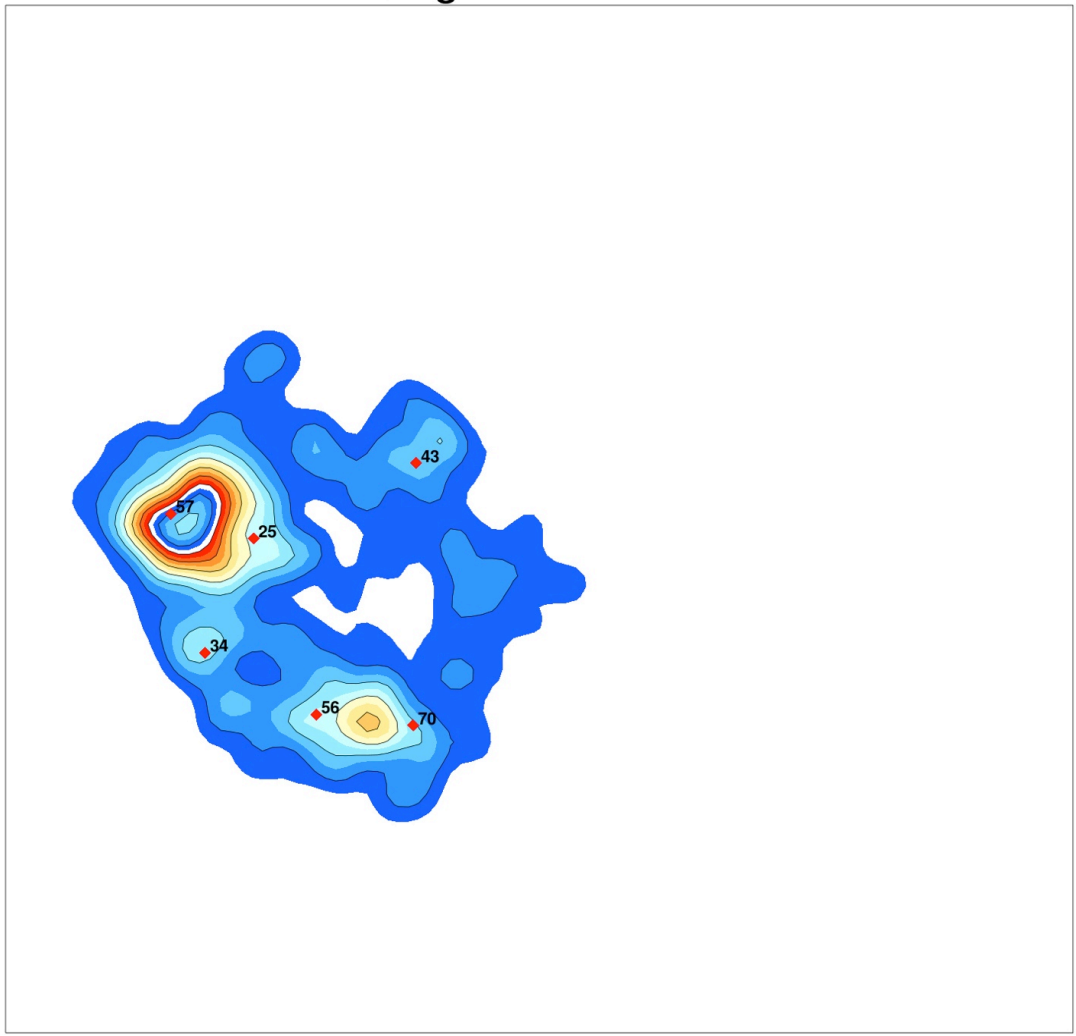

*Figure 7. Normalised average scores for participants 1 to n for each session.*

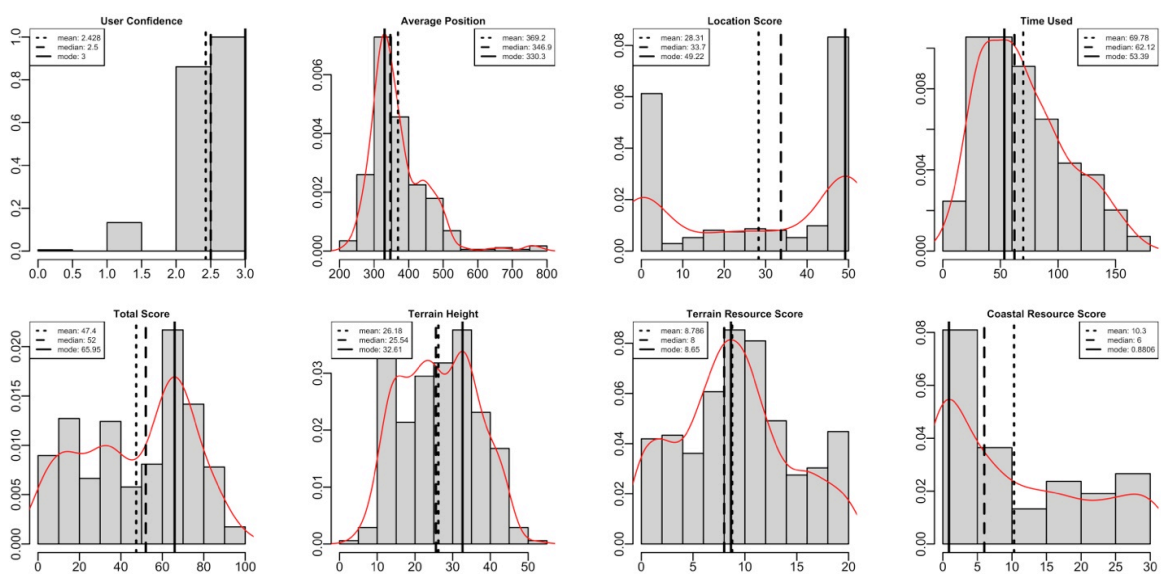

*Figure 8. Histogram showing the statistical distribution of average position of all participants, scores, time used and terrain height of settlements*

The histograms in Figure 8 show the distribution of scores. Coastal proximity scores with bimodal distribution ranged from 0-50 and have two peaks at both extremes of the histogram. The central range is Platykurtic. The majority of participants either obtained high scores from their settlement location or very low scores between 0-10. The majority of participants performed poorly in respect of marine resource scores. The terrain resource scores have a normal distribution with a mean of 8.767. Time used for finding a suitable settlement is positively skewed, with the majority of participants in the range of between 25-100 over 180 seconds. The amount of time used by participants does not contribute to higher scores. On the contrary, it demonstrates a weak negative correlation of R=-0.104 with a statistical significance of p=0.054.

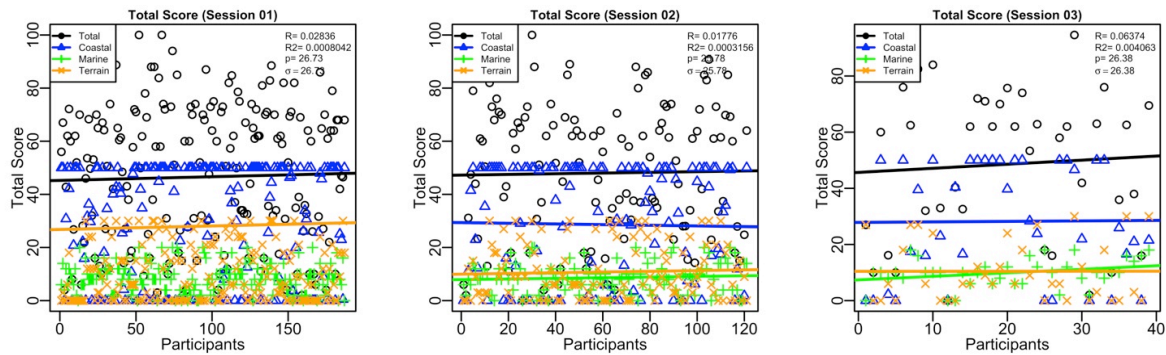

*Figure 9. Looking for evidence of stigmergy across the three simulation sessions*

No statistical significance was found in the sequence of participants in Figure 9. As seen in the regression analysis, there was no evidence that participants learned from previous settlement builders across all sessions of the simulation. However, by correlating the time difference and scores between current and previous settlements, some significance (p<0.05) can be observed even though the relationship is weak. The final two regressions in Figure 10 suggest that participants who observed the behaviour or overheard the conversation of earlier players did learn from their experience. As a result, they have a tendency towards higher scores as compared to participants that did not observe prior participants in the simulation. This is slightly clearer in respect of the terrestrial resource scores rather than the marine resource. The range between 0 to 10 minutes in the Time Difference (*x* axis) between two participants indicates continuity in participation. Participants used the simulation one after the other. The 10 minutes interval between two participants was used for resetting the simulation, preparing users, explanation of concepts prior to starting the simulation and pre-simulation user interface tutorial. The mean of the time difference is 7.05 minutes between two participants, one after the other and for all the sessions involving the 347 users. The weak correlations in the graphs indicate that not all participants learnt from prior settlers via observations, or depended on prior settlers for their decisions to build a settlement. A small number of participants however, did learn from past experience of participants, an indication of stigmergy at work.

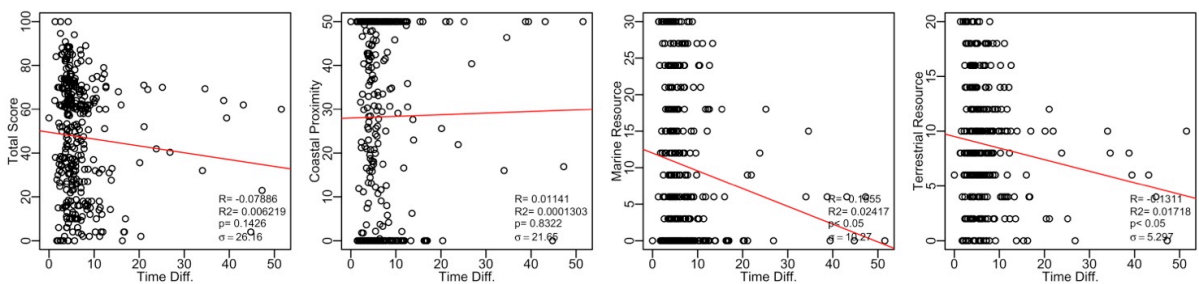

*Figure 10. Regression analysis of time (x) and scores (y). The plot shows observations and learning by a small number of subsequent settlements builders over time, contributing to higher scores as a result. The time difference is measured in minutes.*

## 5. Discussion

The results of the experiment are of considerable interest. The introduction to the paper stressed that although stigmergy is rarely formally explored, or perhaps understood, by archaeologists there are situations in the past in which we might expected stigmergy to play a role. Migration of settlers in the event of widespread catastrophe is a prosaic example - although the evolution of ritual activity and monument construction is an area of very great interest to archaeologists and social scientists that might be usefully explored using models that incorporate the role of stigmergy.

The fact that the participants in this study demonstrate some level of stigmergy using an interactive tabletop suggests that we can begin to use the concept in a fundamental manner to provide appropriate inputs for complex models of the past and to have some confidence in their outputs via virtual environments, agent-based modelling and a suitable interactive display. In doing so we can also begin to explore some of the most intriguing aspects of human culture. How and why did ritual activity begin to evolve formal structures over extremely long periods of time with, apparently, complex and cohesive plans and sophisticated functions? How did migration occur, what was the potential impact of settlers on pre-existing populations and over what period of time did such events unfold? All of these issues have contemporary resonance. Society has not stopped evolving and the role of belief structures is as important today as it was in the distant past. Migration also remains an issue today, as does the effect of climate change. Whilst there may be no direct analogues for the behaviour that we may see or model for the archaeological past the results of such a study will provide a cautionary tale, at least, when we begin to consider the future impact of contemporary climate change or migration on modern populations, the majority of whom live on the coast, much as hunter gathers did 12000 years ago!

**Acknowledgements**

The authors acknowledge the support of the University of Birmingham's Digital Humanities Hub and IBM Visual and Spatial Technology Centre, without which the research would not have been possible. The authors would also like to express appreciation and gratitude to The Royal Society for the privilege extended to the team through the invitation to attend the Royal Society Summer Science Exhibition, held at the Carlton House Terrace, London from 2-7 July 2012.